\documentclass[aps,prl,oneocolumn,amsmath,amssymb,superscriptaddress]{revtex4-1}
\usepackage{graphicx}
\usepackage{amssymb}
\usepackage{natbib}
\usepackage{color}

\newcommand{\ys}[1]{\textcolor{black}{#1}}

\begin{document}


\title{Nature of the spin resonance mode in CeCoIn$_5$}

\author{Yu Song}
\email{yusong@berkeley.edu}
\altaffiliation{Current address: Department of Physics, University of California, Berkeley, California 94720, USA}
\affiliation{Department of Physics and Astronomy, Rice University, Houston, Texas 77005, USA }

\author{Weiyi Wang}
\affiliation{Department of Physics and Astronomy, Rice University, Houston, Texas 77005, USA }

\author{John S. Van Dyke}
\affiliation{Department of Physics, University of Illinois at Chicago, Chicago, Illinois 60607, USA}

\author{Naveen Pouse}
\affiliation{Department of Physics, University of California, San Diego, La Jolla, California 92093, USA}
\affiliation{Center for Advanced Nanoscience, University of California, San Diego, La Jolla, California 92093, USA}

\author{Sheng Ran}
\affiliation{Department of Physics, University of California, San Diego, La Jolla, California 92093, USA}
\affiliation{Center for Advanced Nanoscience, University of California, San Diego, La Jolla, California 92093, USA}

\author{Duygu Yazici}
\affiliation{Department of Physics, University of California, San Diego, La Jolla, California 92093, USA}
\affiliation{Center for Advanced Nanoscience, University of California, San Diego, La Jolla, California 92093, USA}

\author{A. Schneidewind}
\affiliation{J\"{u}lich Center for Neutron Science JCNS, Forschungszentrum J\"{u}lich GmbH, Outstation at MLZ, D-85747, Garching, Germany}

\author{Petr $\rm \check{C}$erm\'{a}k}
\affiliation{J\"{u}lich Center for Neutron Science JCNS, Forschungszentrum J\"{u}lich GmbH, Outstation at MLZ, D-85747, Garching, Germany}
\affiliation{Charles University, Faculty of Mathematics and Physics, Department of Condensed Matter Physics, Ke Karlovu 5, 121 16, Praha, Czech Republic}

\author{Y. Qiu}
\affiliation{NIST center for Neutron Research, National Institute of Standards and Technology, Gaithersburg, Maryland 20899, USA}

\author{M. B. Maple}
\affiliation{Department of Physics, University of California, San Diego, La Jolla, California 92093, USA}
\affiliation{Center for Advanced Nanoscience, University of California, San Diego, La Jolla, California 92093, USA}

\author{Dirk K. Morr}
\email{dkmorr@uic.edu}
\affiliation{Department of Physics, University of Illinois at Chicago, Chicago, Illinois 60607, USA}

\author{Pengcheng Dai}
\email{pdai@rice.edu}
\affiliation{Department of Physics and Astronomy,
Rice University, Houston, Texas 77005, USA }

\begin{abstract}
Spin-fluctuation-mediated unconventional superconductivity can emerge at the border of magnetism, featuring a superconducting order parameter that changes sign in momentum space. Detection of such a sign-change is
experimentally challenging, since most probes are not phase-sensitive. The observation of a spin resonance mode (SRM) from inelastic neutron scattering is often seen as strong phase-sensitive evidence for a sign-changing superconducting order parameter, by assuming the SRM is a spin-excitonic bound state.  
Here, we show that for the
heavy fermion superconductor CeCoIn$_5$, its SRM
defies expectations for
a spin-excitonic bound state, and is not a manifestation of sign-changing superconductivity. 
Instead, the SRM in CeCoIn$_5$ likely arises from a reduction of damping to a magnon-like mode in the superconducting state, due to its proximity to magnetic quantum criticality. 
Our findings emphasize the need for more stringent tests of whether SRMs are spin-excitonic, when using their presence to evidence sign-changing superconductivity.
\end{abstract}

\maketitle


\section{Introduction}

Understanding the physics of unconventional superconductors, which include cuprate, iron-based and heavy fermion superconductors, remains a major challenge in condensed matter physics. Unlike conventional superconductors with phonons responsible for binding electrons into pairs, pairing in unconventional superconductors occurs due to electronic interactions \cite{MNorman,GStewart,BKeimer}. The proximity to magnetically ordered states in these materials suggests spin fluctuations  
as a common thread that can pair electrons in unconventional superconductors \cite{BKeimer,DScalapino,dai}. Unlike phonon-mediated conventional superconductors with superconducting order parameters $\Delta({\bf k})$ that depend weakly on momentum ${\bf k}$, spin-fluctuation-mediated superconductivity requires a $\Delta({\bf k})$ that changes sign in momentum space \cite{DScalapino}. Therefore, the experimental determination of whether a sign-change occurs in $\Delta({\bf k})$ is paramount for identifying and testing the spin-fluctuation-mediated pairing mechanism. 

While sign-changing superconductivity in cuprate superconductors has been confirmed through phases-sensitive tunneling experiments \cite{Harlingen,CCTsuei}, such direct experimental evidence is lacking in most other systems where a sign-change has been proposed. Most experimental techniques, including penetration depth, specific heat, thermal conductivity, and angle-resolved photoemission, can probe the magnitude of the superconducting order parameter and its momentum dependence \cite{CCTsuei}, but are not phase-sensitive. The observation of a spin resonance mode (SRM) in inelastic neutron scattering is commonly regarded as strong phase-sensitive evidence for a sign-changing superconducting order parameter \cite{DScalapino,Eschrig,Mignod,ADChristianson08,CStock08}, based on the assumption that the SRM is a spin-exciton appearing below the particle-hole continuum onset (PHCO), and at a momentum transfer ${\bf Q}$ that connects parts of the Fermi surface exhibiting a sign-change in the superconducting order parameter [$\Delta({\bf k})=-\Delta({\bf k}+{\bf Q})$]. 

Experimentally, the SRM is typically identified through the appearance of 
additional magnetic scattering
in the superconducting state relative to the normal state, peaking at a well-defined energy $E_{\rm r}$ and an intensity that tracks the superconducting order parameter \cite{Eschrig}. While such behaviors of the SRM are consistent with the spin-exciton scenario, alternative explanations have also been proposed \cite{Eschrig,DKMorr,FKruger2007,GXu2009,SOnari1,AVChubukov}. Moreover, phenomenologically similar enhanced scattering in the superconducting state have been observed in systems without sign-changing superconductivity, including phonons \cite{HKawano1996,CStassis1997} and hydrogen tunneling excitations \cite{AMagerl} in conventional superconductors and the resonant magnetic exciton mode in semiconducting rare-earth borides \cite{GFriemel2012,KSNemkovski}, indicating mechanisms other than sign-changing superconductivity that could account for the experimental signatures of the SRM. Therefore, it is important to test whether experimentally observed SRMs are indeed spin-excitonic in nature, given the presence of a SRM is often used to evidence sign-changing unconventional superconductivity. This is underscored by recent measurements on CeCu$_2$Si$_2$ that demonstrated it exhibits nodeless superconductivity \cite{Kittaka,TYamashita,TTakenaka,GPang2018,YLi2018}, despite the observation of a SRM which suggests nodal $d$-wave superconductivity in the spin-exciton scenario \cite{OStockert,IEremin}. 


In this work we use inelastic neutron scattering to systematically study the SRM in the prototypical heavy fermion superconductor CeCoIn$_5$ ($T_{\rm c}=2.3$ K) 
\cite{Thompson,maple}, which exhibits sign-changing $d_{x^2-y^2}$-wave superconductivity similar to the cuprates \cite{KIzawa,BBZhou,MPAllan13}. 
\ys{Contrary to expectations for a spin-excitonic SRM with a prominent downward dispersion, our results show that the SRM in CeCoIn$_5$ disperses upward without downward-dispersing features.
Under applied magnetic field, the SRM splits into two upward-dispersing branches, with the dispersive features becoming progressively smeared out due to an increase in damping. Taken together, our results suggest that the SRM in CeCoIn$_5$ is not spin-excitonic, and therefore is not a manifestation of the $d_{x^2-y^2}$-wave superconducting order parameter. Instead, it likely results from the removal of damping to a pre-existing magnetic mode in a more strongly coupled unconventional superconductor.
Our findings underscore the importance of more stringent tests to verify the spin-excitonic nature of SRMs, when using their presence to evidence sign-changing unconventional superconductivity.}

\begin{figure}[t]
	\includegraphics[scale=.5]{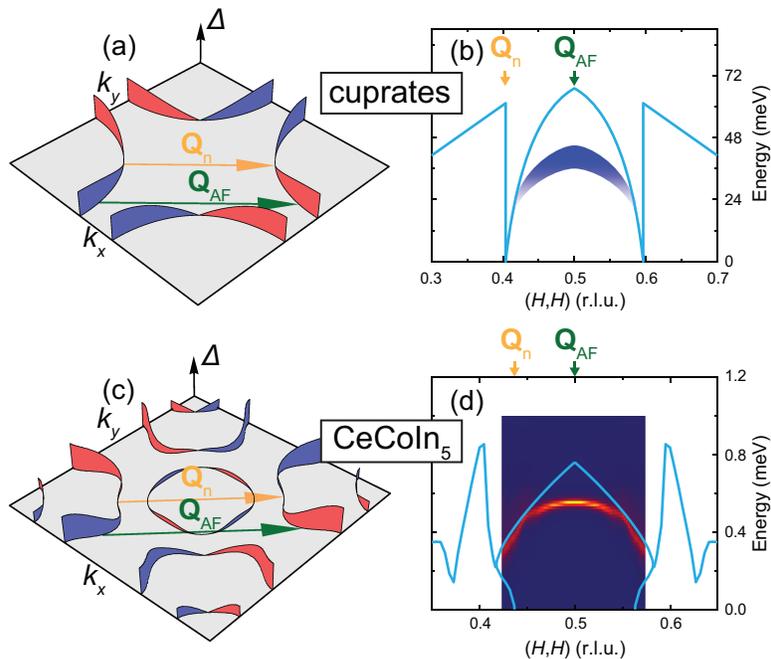}
	\caption{
		\textbf{The spin resonance mode (SRM) in the spin-exciton scenario.} (a) Fermi surface and the $d_{x^2-y^2}$-wave superconducting order parameter in the cuprates \cite{IEremin2005}. (b) Schematic dispersion of the SRM in the cuprates along the $(H,H)$ direction. The SRM in the cuprates falls below the particle-hole continuum onset (PHCO), indicated by the light blue lines. (c) Fermi surfaces and the $d_{x^2-y^2}$-wave superconducting order parameter in CeCoIn$_5$ \cite{JVanDyke2014}. (d) Calculated dispersion of the SRM in CeCoIn$_5$, in the spin-exciton scenario (see Supplementary Note 1 for details). The light blue lines indicate the PHCO. The red and blue surfaces in (a) and (c) represent superconducting order parameters with opposite signs. ${\bf Q}_{\rm AF}=(0.5,0.5)$ connects hot spots on the Fermi surface that exhibits a robust superconducting gap, ${\bf Q}_{\rm n}$ connects parts of the Fermi surface that correspond to nodes of the superconducting order parameter.
	}
\end{figure}

\section{Results}
\subsection{Dispersion of the SRM in CeCoIn$_5$ at zero-field}
In the spin-exciton scenario, the SRM is a bound state residing below the PHCO with $E({\bf Q})<\rm{min}(\mid\Delta({\bf k})\mid+\mid\Delta({\bf k}+{\bf Q})\mid)$, resulting from a sign-change in the superconducting order parameter \cite{DScalapino,Eschrig}. For cuprates with a $d_{x^2-y^2}$-wave superconducting order parameter, the SRM peaks at the antiferromagnetic wavevector ${\bf Q}_{\rm AF}=(0.5,0.5)$, which connects hot spots that are close to the antinodal points of the $d_{x^2-y^2}$-wave superconducting order parameter [Fig. 1(a)]. As the SRM disperses away from ${\bf Q}_{\rm AF}$ towards ${\bf Q}_{\rm n}$, which connects the nodal points of the superconductivity order parameter, the PHCO is progressively pushed towards zero. The reduction of the PHCO away from ${\bf Q}_{\rm AF}$ requires a spin-excitonic SRM to exhibit a downward dispersion away from ${\bf Q}_{\rm AF}$, so that it stays below the PHCO. Inelastic neutron scattering measurements of the SRM in \ys{hole-doped} cuprates demonstrated that it dominantly disperses downwards, consistent with expectations of the spin-exciton picture [Fig. 1(b)] \cite{PBourges2000,dai00,SPailhes2004,Reznik04,Hayden04,Sidis}. For iron pnictide superconductors with isotropic $s^\pm$-wave superconducting gaps, the SRM is also consistent with being a spin-exciton \cite{DScalapino,dai}.  Here, unlike the cuprates, the PHCO depends weakly on momentum ${\bf Q}$, allowing spin-excitonic SRMs to exhibit upward dispersions, as observed in electron- \cite{Kim13} and hole-doped compounds \cite{RZhang}.

In the prototypical heavy fermion superconductor CeCoIn$_5$, like the cuprates, the superconducting order parameter is $d_{x^2-y^2}$-wave \cite{KIzawa,BBZhou,MPAllan13} and the SRM peaks around ${\bf Q}_{\rm AF}$ in momentum and $E_{\rm r}\approx0.6$ meV in energy \cite{CStock08}; thus also like the cuprates, the PHCO is gradually suppressed moving from ${\bf Q}_{\rm AF}$ towards ${\bf Q}_{\rm n}$ [Figs. 1(c) and (d)], resulting in a downward dispersion of the SRM in the spin-exciton scenario [Fig. 1(d), see Supplementary Note 1 for details] \cite{IEremin}. Experimentally, however, the SRM is found to be dominated by a robust upward dispersion for $E\gtrsim E_{\rm r}$, contrary to expectations in the spin-exciton picture \cite{YSong2016,SRaymond2015}. These upward-dispersing features and the strong $L$-dependence of the SRM in CeCoIn$_5$ suggest it is a magnon-like mode, rather than a spin-exciton \cite{YSong2016,AVChubukov}. While the SRM in CeCoIn$_5$ is dominated by an upward-dispersing branch for $E\gtrsim E_{\rm r}$, whether a downward-dispersing branch expected in the spin-exciton scenario also exists for $E<E_{\rm r}$, remains unclear.

\begin{figure}[t]
	\includegraphics[scale=.5]{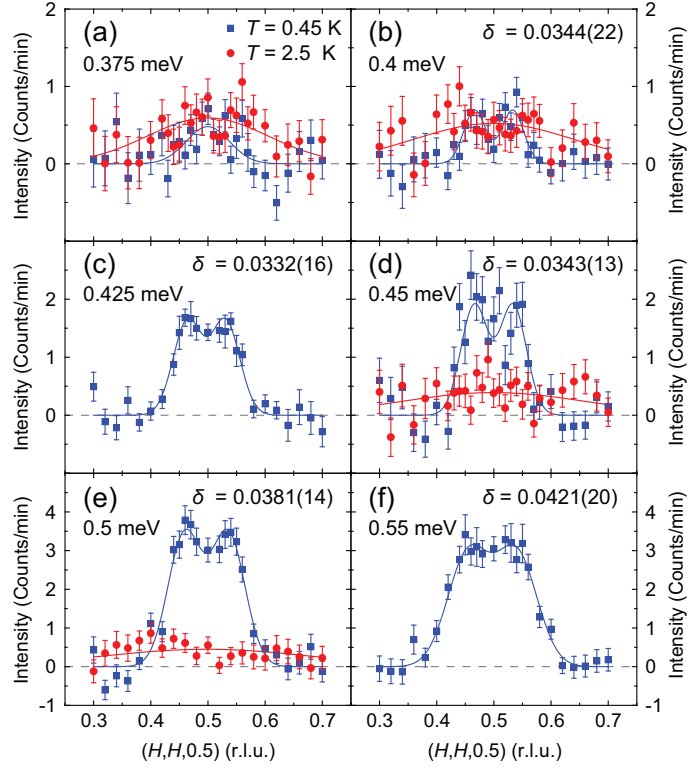}
	\caption{
		\textbf{Constant-energy scans along $(H,H,0.5)$ for $E\lesssim E_{\rm r}$.} Background-subtracted constant-energy scans measured using PANDA, for (a) $E=0.375$ meV, (b) $E=0.4$ meV, (c) $E=0.425$ meV, (d) $E=0.45$ meV, (e) $E=0.5$ meV and (f) $E=0.55$ meV. Blues squares are data at $T=0.45$ K, well below $T_{\rm c}=2.3$ K. Red circles are data at $T=2.5$ K, just above $T_{\rm c}$. Solid blue lines are fits to two Gaussian peaks \ys{centered at $(0.5\pm\delta,0.5\pm\delta,0.5)$} for data in the superconducting state, except for $E=0.375$~meV, which is fit to a single Gaussian peak. Solid red lines are fits to a single Gaussian peak for data in the normal state. A linear background included in the fitting has been subtracted. \ys{For panels (b)-(f), the fit values and uncertainties of $\delta$ are shown in the upper right corner. All vertical error bars in the figures represent statistical errors of 1 s.d.}
	}
\end{figure}

To elucidate whether the SRM in CeCoIn$_5$ has any downward-dispersing features, we carried out detailed inelastic neutron scattering measurements of the SRM in CeCoIn$_5$ using PANDA, along the $(H,H,0.5)$ direction for $E\lesssim E_{\rm r}\approx0.6$ meV, with results shown in Fig. 2. The magnetic scattering at $E=0.375$ meV is weaker in the superconducting state compared to the normal state [Fig. 2(a)], demonstrating a partial gapping of the magnetic fluctuations at this energy upon entering the superconducting state. With increasing energy, scattering in the superconducting state becomes more intense compared to the normal state [Figs. 2(b)-(f)], and the SRM can be clearly identified by such enhanced magnetic scattering. Constant-energy scans along $(H,H,0.5)$ for $E\geq0.4$ meV [Figs. 2(b)-(f)] clearly reveal two peaks at ${\bf Q}=(0.5\pm\delta,0.5\pm\delta,0.5)$, in good agreement with previous work (see Supplementary Fig. 1 and Supplementary Note 2 for details) \cite{SRaymond2015}. While the magnetic scattering for $E=0.375$ meV appears to be a single peak, its broad width compared to higher energies suggests the magnetic scattering at this energy also consists of two peaks. By fitting the results in Figs. 2(b)-(f) using two Gaussian peaks at ${\bf Q}=(0.5\pm\delta,0.5\pm\delta,0.5)$, we find \ys{$\delta$ does not change significantly for $E\leq0.45$~meV [Fig. 2(b)-(d)] and} increases monotonically with increasing energy \ys{for $E>0.45$~meV} [Fig. 3(a)], ruling out any downward-dispersing features. Combined with similar measurements for $E\gtrsim E_{\rm r}$ obtained using Multi-Axis Crystal Spectrometer (MACS) (see Supplementary Figs. 2-3 and Supplementary Note 3 for details), we find the SRM in CeCoIn$_5$ disperses only upward, inconsistent with calculations for the spin-exciton scenario, based on an electronic structure from scanning tunneling microscopy measurements [Figs. 1(d) and 3(a), see Supplementary Note 1 for details] \cite{MPAllan13,JVanDyke2014}. Instead, the dispersion of the SRM resemble spin waves in CeRhIn$_5$ [Fig. 3(b)] \cite{PDas2014,CStock2015}, suggesting it to be magnon-like (see Supplementary Fig. 4 and Supplementary Note 2 for additional comparisons).

\begin{figure}[t]
	\includegraphics[scale=.6]{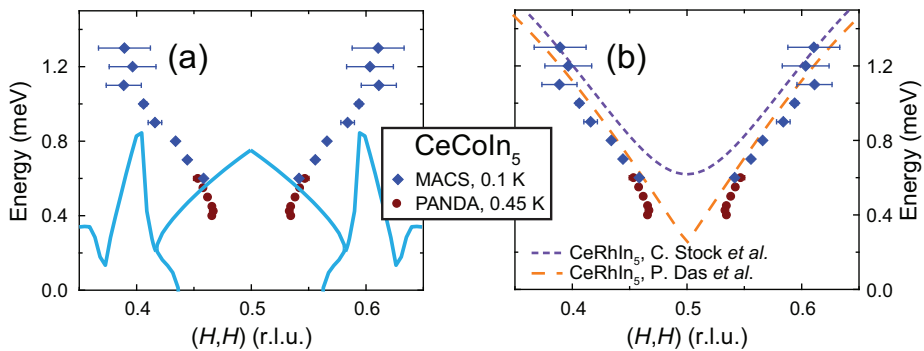}
	\caption{
		\textbf{Zero-field dispersion of the spin resonance mode (SRM).} (a) The experimentally observed dispersion of the SRM in CeCoIn$_5$, compared with the particle-hole continuum onset (PHCO) (light blue lines). (b) The experimentally observed dispersion of the SRM in CeCoIn$_5$, compared with spin waves in CeRhIn$_5$ \cite{PDas2014,CStock2015}. Horizontal error bars are least-square fit errors (1 s.d.), \ys{diamond} symbols are from Multi-Axis Crystal Spectrometer (MACS) data (see Supplementary Note 3 for details) and \ys{circle} symbols are from PANDA data (Fig. 2). 
	}
\end{figure}

\subsection{Splitting of the dispersive SRM under applied magnetic field}


For a spin-excitonic SRM that is isotropic in spin space, the application of a magnetic field should split it into a triplet in energy \cite{JPIsmer}. 
In CeCoIn$_5$, the application of an in-plane magnetic field splits the SRM into a doublet, rather than a triplet \cite{CStock2012,SRaymond2012}, likely due to the presence of magnetic anisotropy \cite{AAkbari,YSong2016}.
The doublet splitting of the SRM under applied field, combined with the upward dispersion, raises the question of how the dispersive features of the SRM in CeCoIn$_5$ evolve with applied field, and whether the absence of a downward-dispersing branch is robust under applied field. 

To address these questions, we studied the SRM in CeCoIn$_5$ using MACS, under an applied magnetic field perpendicular to the $[H,H,L]$ scattering plane, with results shown in Figs. 4 and 5. Constant-energy scans along $(H,H,0.5)$ and $(0.5,0.5,L)$ directions in Fig. 4 reveal dramatic changes to the SRM away from ${\bf Q}_{\rm AF}$ under applied magnetic field. For $E=0.5$ meV and $E=0.6$ meV, the SRM broadens upon increasing the magnetic field from $B=0$ T to $B=6$ T [Figs. 4(a)-(d)]. On the other hand, for $E=0.8$ meV and $E=1.0$ meV, two split peaks around ${\bf Q}_{\rm AF}$ are clearly seen at $B=0$ T, while increasing the magnetic field to $B=3$~T significantly reduces the splitting and only a single peak can be resolved  at $B=6$ T [Figs. 4(e)-(h)]. 
We note that while the SRM is peaked slightly away from ${\bf Q}_{\rm AF}$ for $E\lesssim E_{\rm r}$ at zero-field, as demonstrated in Fig. 2, the resolution of our MACS measurements is insufficient to resolve such a small splitting, instead a single peak at ${\bf Q}_{\rm AF}$ is observed [Figs. 4(a) and (c)].

These disparate behaviors at different energies can be understood to result from the doublet splitting of the upward-dispersing SRM, as schematically depicted in Fig. 4(i). 
The broadening of the peaks along $(H,H,0.5)$ at $E=0.5$ and 0.6~meV under applied field
is due to a downward shift of the lower branch of the SRM, and increased damping resulting from the PHCO also moving to lower energies.
For higher energies $E=0.8$ and 1.0 meV, the intensity of magnetic scattering is dominated by the upper branch, and because the upper branch moves to higher energies under applied field, a reduction in peak splitting is observed. Our results indicate the dispersive SRM in CeCoIn$_5$ splits into two branches under an in-plane magnetic field, while maintaining its upward-dispersing character. This conclusion is also supported by the analysis of peak splittings for the data in Fig. 5 (See Supplementary Fig. 5 and Supplementary Note 4 for details).

In addition to splitting the SRM into two upward-dispersing branches, energy-$(H,H,0.5)$ and energy-$(0.5,0.5,L)$ maps in Fig. 5 and Supplementary Fig. 6 (\ys{obtained from data shown in Supplementary Figs. 3,7 and 8}, see Supplementary Note 3 for details) suggest that applied magnetic field also results in significant damping to the SRM in CeCoIn$_5$ (see Supplementary Fig. 9 and Supplementary Note 4 for additional evidence from constant-${\bf Q}$ scans). While the dispersive features can be clearly observed in the $B=0$ T data [Figs. 5(a) and (b)], with applied field the dispersive features become less prominent for $B=4$ T [Figs. 5(c) and (d)] and  for $B=6$ T no dispersive features can be resolved [Supplementary Figs. 6(d) and (h)]. These results suggest that with applied field, the SRM becomes progressively damped and its dispersive character smeared out, becoming similar to overdamped magnetic excitations in the normal state, as the applied field approaches the upper critical field (see Supplementary Fig. 10 and Supplementary Note 3 for details). \ys{The increase in damping is unexpected in the spin-exciton scenario. This is because the SRM and the PHCO are shifted in energy in unison by an applied magnetic field, the SRM should therefore remain undamped (see Supplementary Figure 11 and Supplementary Note 1 for details). Instead, the observed damping of the SRM with increasing field suggests that the SRM and the PHCO move independently with increasing magnetic field, consistent with the suggestion that the SRM in CeCoIn$_5$ results from the removal of damping to a pre-existing magnetic mode in the superconducting state \cite{AVChubukov,YSong2016}, rather than being a spin-exciton.}

\begin{figure}[t]
	\includegraphics[scale=.5]{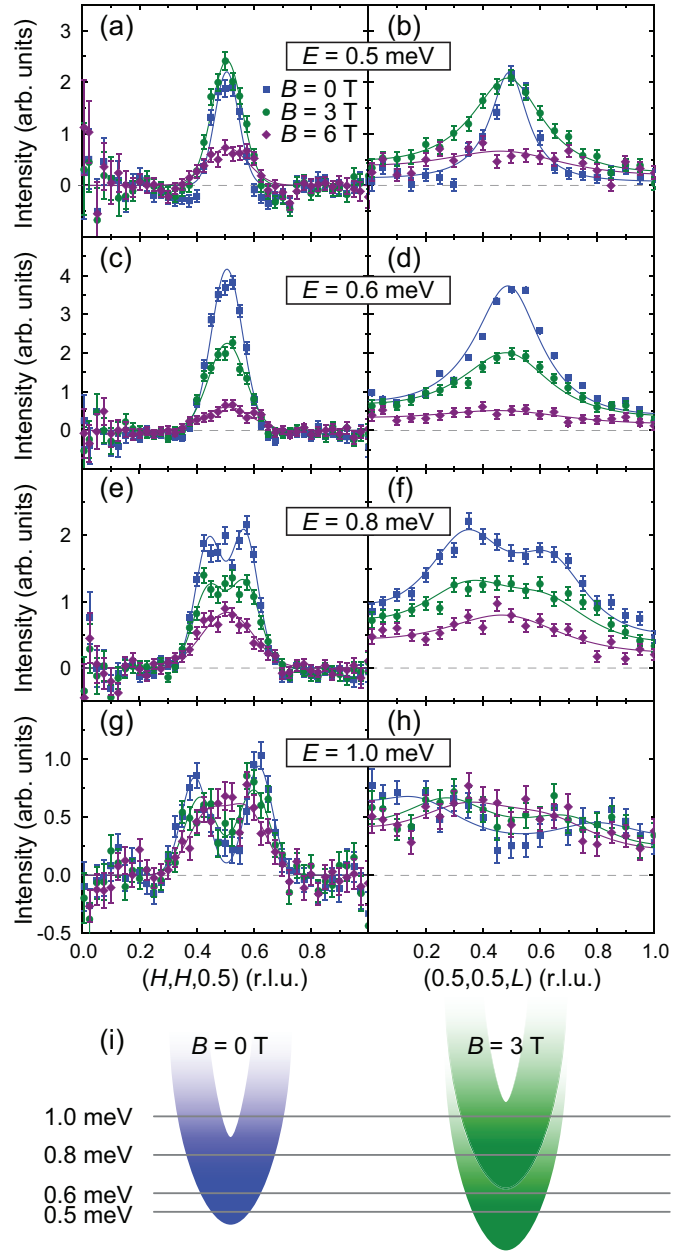}
	\caption{
		\textbf{Constant-energy scans under applied magnetic field.} 
		Constant-energy scans along $(H,H,0.5)$ measured using Multi-Axis Crystal Spectrometer (MACS), for (a) $E=05$ meV, (c) $E=0.6$ meV, (e) $E=0.8$ meV and (g) $E=1.0$ meV, under different applied magnetic fields. Constant-energy scans along $(0.5,0.5,L)$, for (b) $E=0.5$ meV, (d) $E=0.6$ meV, (f) $E=0.8$ meV and (h) $E=1.0$ meV. The normal state magnetic scattering measured at $T=2.5$ K has been subtracted. For $(H,H,0.5)$ scans, the solid lines are fits to one or two Gaussian peaks; for $(0.5,0.5,L)$ scans, the solid lines are fits to a lattice sum of one or two Lorentzian peaks. All vertical error bars in the figure represent statistical errors of 1 s.d. (i) Schematic doublet splitting of the spin resonance mode (SRM) in CeCoIn$_5$ under applied magnetic field, resulting in two branches that both disperse upward.
	}
\end{figure}



\begin{figure}[t]
	\includegraphics[scale=0.5]{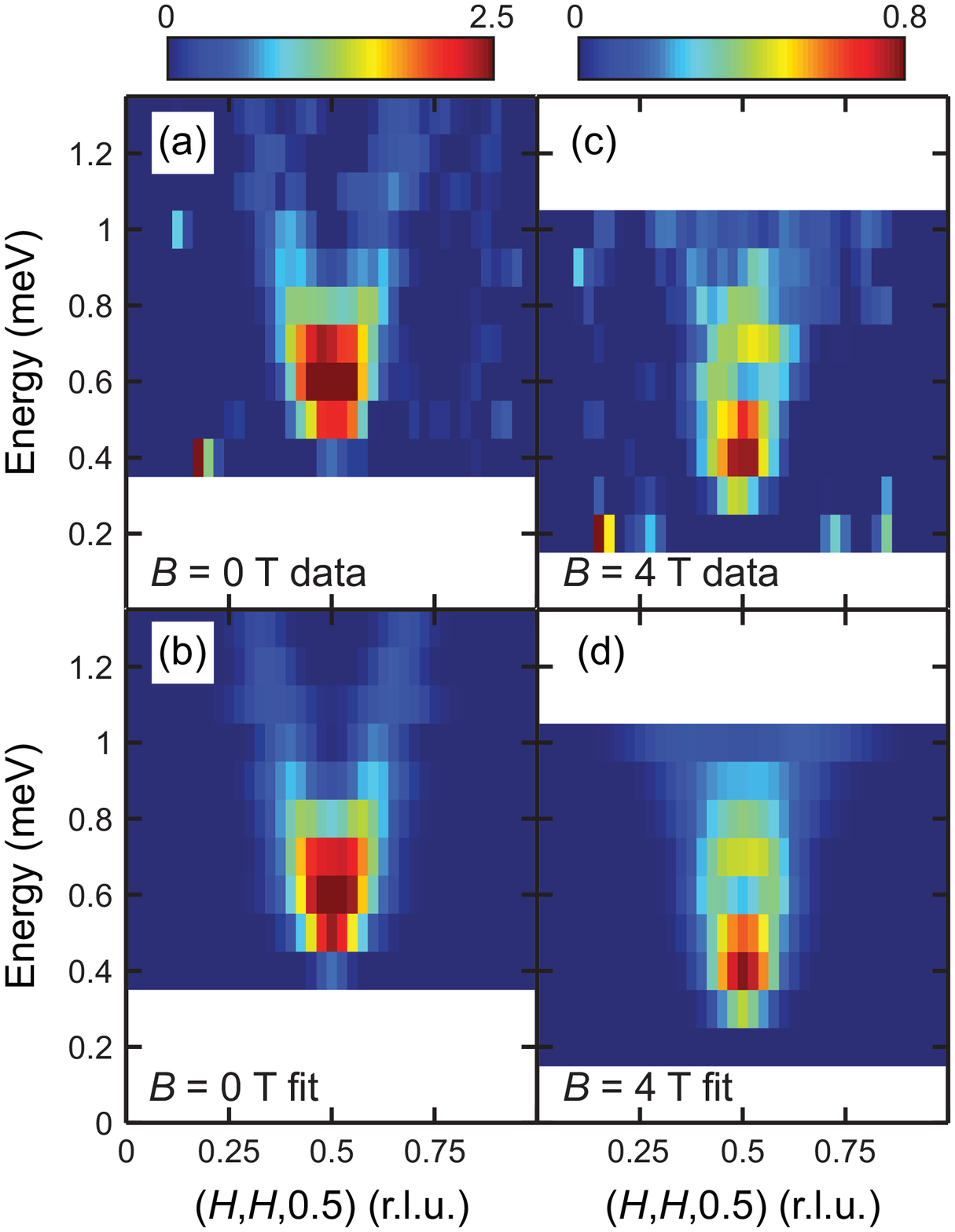}
	\caption{
		\textbf{Energy-$(H,H,0.5)$ maps under applied magnetic field.}
		Energy-$(H,H,0.5)$ maps of the spin resonance mode (SRM) in CeCoIn$_5$ measured using Multi-Axis Crystal Spectrometer (MACS), for (a) $B=0$ T, and (b) $B=4$ T, with the corresponding fits shown in (c) and (d), respectively. The normal state magnetic scattering measured at $T=2.5$ K has been subtracted. $B=0$ data are measured with $E_{\rm f}=3.7$ meV, and $B=4$ T data are measured with $E_{\rm f}=3.0$ meV. The fits are obtained by combining individual fits to line cuts (see Supplementary Note 3 for details).
	}
\end{figure}

\section{Discussion}

Our results demonstrate the SRM in CeCoIn$_5$ disperses upward, without downward dispersing features, inconsistent with expectations for a spin-exciton in a $d_{x^2-y^2}$-wave superconductor. This suggests that either the superconducting order parameter in CeCoIn$_5$ is not $d_{x^2-y^2}$-wave, or that the SRM is not spin-excitonic. While nodeless $s^{\pm}$ superconductivity has been proposed for Pu-based 115 heavy-fermion superconductors \cite{FRonning2012,TDas2015}, there is strong experimental evidence for $d_{x^2-y^2}$-wave superconductivity in CeCoIn$_5$ with a robust nodal $d_{x^2-y^2}$-wave superconducting order parameter \cite{KIzawa,BBZhou,MPAllan13,YSong2016,YXu2016}. Therefore, our findings indicate the SRM in CeCoIn$_5$ is not spin-excitonic in origin, and as such, it is not a manifestation of the sign-changing $d_{x^2-y^2}$-wave
superconducting order parameter in CeCoIn$_5$. More broadly, our results highlight that while SRMs in different unconventional superconductors exhibit similar experimental signatures, they may have distinct origins. When a SRM is spin-excitonic in origin, it evidences sign-changing superconductivity and provides information about the system's electronic structure. On the other hand, if the SRM has a different origin, it may not be appropriate to use the observation of a SRM for these purposes. \ys{We note that while a spin-excitonic contribution to the SRM with intensity weaker than our detection limit cannot be ruled out, this does not affect our conclusion that the detectable SRM in CeCoIn$_5$ is not spin-excitonic.} 

In the cuprates, X- or Y-shaped excitations with dominant upward dispersing branches, which may result from either localized or itinerant electrons, have been observed \cite{Sidis,JMTranquada2004,Fujita_JPSJ,VHinkov2007,MKChan_NC_2016,MKChan_PRL_2016}. However, these upward-dispersing excitations are different from what we have observed in CeCoIn$_5$ in that they are already present in the normal state, and exhibits little or no change upon entering the superconducting state, i.e. they are not SRMs. When a SRM is present, as seen through additional magnetic scattering uniquely associated with the superconducting state, it is always dominated by a downward dispersion, as shown in Fig. 1(b). While a weaker upward-dispersing branch of the SRM has also been detected in some cuprates \cite{SPailhes2004,Reznik04,Hayden04}, these SRMs were shown to be consistent with spin-excitons residing in a different region of momentum space \cite{SPailhes2004,IEremin2005}, where the PHCO energy is large [region with $\mid{\bf Q}\mid<\mid{\bf Q}_{\rm n}\mid$ in Fig. 1(b)]. In CeCoIn$_5$, based on the electronic structure extracted from scanning tunneling microscopy measurements \cite{MPAllan13,JVanDyke2014}, it can be seen that while a similar region of the PHCO is present [Fig. 3(a)], it does not account for our experimentally determined dispersion in CeCoIn$_5$. While the observation of a spin-excitonic SRM indicates sign-changing superconductivity, the absence of a SRM (as seen experimentally in sufficiently underdoped cuprates \cite{MKChan_NC_2016}) or the observed SRM not being spin-excitonic (as in CeCoIn$_5$) does not invalidate sign-changing superconductivity, but simply means that in these cases information on the superconducting order parameter do not directly manifest in magnetic excitations.

In addition to demonstrating the SRM in CeCoIn$_5$ disperses upward at zero-field, our results in Fig. 2 also show that the SRM is appropriately described by two peaks at ${\bf Q}=(0.5\pm\delta,0.5\pm\delta,0.5)$ for all energies. The splitting of the SRM for $E<E_{\rm r}$ is suggested to evidence that the SRM is a precursor \cite{SRaymond2015,CStock2012,VPMichal2011} to the field-induced spin-density-wave phase ($Q$-phase) that orders at ${\bf Q}=(0.5\pm\delta_{Q},0.5\pm\delta_{Q},0.5)$ \cite{Kenzelmann08,SGerber2013}. For $E\lesssim0.45$ meV, our data in Fig. 2 shows that $\delta\approx0.034$, significantly smaller than $\delta_Q=0.05$. While CeCoIn$_5$ is magnetically disordered, it can be tuned towards commensurate magnetic order at ${\bf Q}_{\rm AF}$ through Cd- \cite{MNicklas2007}, Rh- \cite{MYokoyama2008} or Hg-doping \cite{CStock2018}, as well as incommensurate magnetic order at ${\bf Q}=(0.5\pm\delta_{Q},0.5\pm\delta_{Q},0.5)$ through Nd-doping \cite{SRaymond_JPSJ} or applying magnetic field \cite{Kenzelmann08}. The proximity of CeCoIn$_5$ to two types of magnetic orders indicates that fluctuations associated with both may be present in CeCoIn$_5$, also suggested by two types of fluctuations unveiled by half-polarized neutron scattering experiments \cite{SRaymond2012}. In such a scenario, the overlap of fluctuations at ${\bf Q}=(0.5\pm\delta_Q,0.5\pm\delta_Q,0.5)$ and ${\bf Q}_{\rm AF}$ results in the observed $\delta<\delta_Q$ for $E\lesssim E_{\rm r}$. Such a coexistence of two types of magnetic fluctuations has also been observed FeTe \cite{YSong2018,DTam2019}; in both CeCoIn$_5$ and FeTe, it results from the quasi-degeneracy of different magnetic states.

While a SRM that is isotropic in spin space is expected to split into triplets in energy under applied field \cite{JPIsmer}, when an easy-plane magnetic anisotropy perpendicular to the field direction $(1\bar{1}0)$ is taken into consideration, it is possible to account for the doublet splitting in CeCoIn$_5$ \cite{AAkbari,YSong2016}. 
However, the SRM has been demonstrated to have an Ising character at zero-field, with the easy-axis along $(001)$ \cite{SRaymond2015}; therefore, to account for the doublet splitting of the SRM in CeCoIn$_5$, it is necessary to consider modifications to the form of magnetic anisotropy under applied magnetic field, \ys{as demonstrated for magnetically ordered CeRhIn$_5$ \cite{DFobes2018}. 
At zero-field, CeRhIn$_5$ exhibits an easy-$ab$-plane spin anisotropy, an applied magnetic field along $(1\bar{1}0)$ induces an anomalously large additional easy-axis anisotropy along $(110)$, driving the system to exhibit an easy-axis spin anisotropy along $(110)$ overall. 
In the case of CeCoIn$_5$, at zero-field it exhibits an easy-axis spin anisotropy along $(001)$, and if an applied magnetic field along $(1\bar{1}0)$ also eases the spin anisotropy along $(110)$ as in CeRhIn$_5$, the system may be driven to overall exhibit an easy-plane-like spin anisotropy, with the easy-plane spanned by $(001)$ and $(110)$ (perpendicular to the applied field).}

In conclusion, our detailed inelastic neutron scattering measurements indicate the SRM in CeCoIn$_5$ disperses upward without any downward dispersing features, indicating it is not spin-excitonic in origin. Under an applied magnetic field, the SRM splits into two upward-dispersing branches and progressively loses its dispersive characters with increasing field, suggesting the SRM in CeCoIn$5$ results from the removal of damping to a pre-existing magnetic mode in the superconducting state. As such, our results suggest the SRM in CeCoIn$_5$ is not a result of the sign-change in its superconducting order parameter. Our findings demonstrate SRMs observed in unconventional superconductors can have origins other than spin-excitonic, in which case their presence may not provide information on the superconducting order parameter. 

\section{Methods}
\subsection{Sample preparation and neutron scattering experimental setups}
Single crystals of CeCoIn$_5$ were prepared by the indium self-flux method \cite{VSZapf}. Hundreds of CeCoIn$_5$ single crystals with a total mass $\sim$1 g were co-aligned in the $[H,H,L]$ scattering plane on aluminum plates using a hydrogen-free glue. Magnetic field is applied perpendicular to the scattering plane, along the $(1\bar{1}0)$ direction.

Neutron scattering experiments were carried out on the PANDA cold three-axes spectrometer at the Heinz Maier-Leibnitz Zentrum \cite{panda} and the Multi-Axis Crystal Spectrometer (MACS) at the NIST Center for Neutron Research. The inelastic neutron scattering experiments on PANDA used fixed $k_{\rm f}=1.3$ \AA$^{-1}$. A sapphire filter is used before the monochromator and a Be filter cooled to 40~K is used before the sample. The monochromator has horizontal and vertical variable focusing mechanics, vertical focusing of the analyzer is fixed (variable focusing is not needed because the detector is a vertically placed 1 inch $^{3}$He tube) and horizontal focusing is variable. In the focused mode, variable focusings are adjusted depending on the neutron wavelength based on empirically optimized values. The inelastic neutron scattering measurements at MACS used Be filters both before and after the sample with fixed $E_{\rm f}=3.0$~meV or $E_{\rm f}=3.7$~meV. Most of measurements on MACS were made using the 20 spectroscopic detectors simultaneously to efficiently obtain the magnetic scattering within the $[H,H,L]$ scattering plane. Constant-${\bf Q}$ scans at ${\bf Q}_{\rm AF}$ Supplementary Fig. 9 were carried out using MACS with a single detector. The analyzers are vertically focused, while the monochromator is doubly focused.

\subsection{Data analysis}
Data shown in Fig. 2 and Supplementary Fig. 1 are obtained using PANDA. The constant-energy scans were fit to a single Gaussian peak or two Gaussian peaks equally displaced from the center; scans at different energy transfers are fit globally with the same peak center. Constant-${\bf Q}$ scans in Supplementary Fig. 9 are measured on MACS using a single detector. All the rest of neutron scattering data are obtained using MACS by measuring maps of large portions of the $[H,H,L]$ scattering plane, simultaneously using the 20 detectors available at MACS. The maps of $[H,H,L]$-plane are folded into a single quadrant to improve statistics. Cuts along $(H,H,0.5)$ were obtained by binning data with $0.37\leq L\leq0.63$ and a step size of 0.025; cuts along $(0.5,0.5,L)$ are obtained by binning data with $(0.42,0.42)\leq (H,H)\leq (0.58,0.58)$ and a step size of 0.05. Normal state magnetic excitations measured at $T=2.5$ K have been subtracted in all the MACS data except Supplementary Fig. 10. The cuts along $(H,H,0.5)$ are fit with a single Gaussian peak centered at ${\bf Q}=(0.5,0.5,0.5)$ or two Gaussian peaks at ${\bf Q}=(0.5\pm\delta,0.5\pm\delta,0.5)$. The cuts along $(0.5,0.5,L)$ are fit using a lattice sum of a single Lorentzian peak centered at ${\bf Q}=(0.5,0.5,0.5)$ or a lattice sum of two Lorentzian peaks at ${\bf Q}=(0.5,0.5,0.5\pm\delta)$. $B=4$ T data are collected using $E_{\rm f}=3.0$ meV, while measurements at other fields used $E_{\rm f}=3.7$ meV. 
Using MACS we collected high statistics data for selected energies (Fig. 4 and Supplementary Fig. 2) and lower statistics data with finer energy steps (Fig. 5 and Supplementary Figs. 3, 6-8). The zero-field data shown in Fig. 4 and Supplementary Fig. 2 are reproduced from Ref. \cite{YSong2016}, to compare with data under applied field.   

{\bf DATA AVAILABILITY:} 

All relevant data are available from the corresponding authors upon reasonable request.

{\bf ACKNOWLEDGEMENTS:} 

We thank S. Raymond and C. Stock for helpful discussions. The neutron scattering work at Rice is supported by the U.S. DOE, BES under grant no. DE-SC0012311 (P.D.). Part of the material characterization efforts at Rice is supported by the Robert A. Welch Foundation Grant Nos. C-1839 (P.D.). Research at UC San Diego was supported by the US Department of Energy, Office of Basic Energy Sciences, Division of Materials Sciences and Engineering, under Grant No. DEFG02-04-ER46105 (single crystal growth) and US National Science Foundation under Grant No. DMR-1810310 (characterization of physical properties). Access to MACS was provided by the Center for High Resolution Neutron Scattering, a partnership between the National Institute of Standards and Technology and the National Science Foundation under Agreement No. DMR-1508249. 

{\bf AUTHOR CONTRIBUTIONS:}

Y. S. and P. D. led the project. The neutron scattering experiments were
performed by Y. S., W. W., A. S., P. C. and Y. Q. The samples were prepared by N. P., S. R., D. Y. and M. B. M.
Y. S. co-aligned the samples. Y. S. and W. W. analyzed the data. Theoretical calculations were carried out by J. V. and D. K. M. The manuscript was written by Y. S., D. K. M. and P. D. with input from all coauthors. 

{\bf COMPETING INTERESTS:} 

The authors declare no competing interests.

\ys{Correspondence and
requests for materials should be addressed to Y. S. (yusong@berkeley.edu), D. K. M. (dkmorr@uic.edu) or P. D. (pdai@rice.edu).}


\end{document}